\documentclass{midl} 

\usepackage{mwe} 
\jmlrproceedings{MIDL}{Medical Imaging with Deep Learning}
\jmlrpages{}
\jmlryear{2019}
\jmlrworkshop{MIDL 2019 -- Extended Abstract Track}
\usepackage{bm}
\usepackage[multi-part-units=single]{siunitx}
\title[4D Spatio-Temporal Deep Learning with 4D fMRI Data]{4D Spatio-Temporal Deep Learning with 4D fMRI Data for Autism Spectrum Disorder Classification}

\midlauthor{\Name{Marcel Bengs\midljointauthortext{Contributed equally}\nametag{$^{1}$}} \Email{marcel.bengs@tuhh.de}\\
\Name{Nils Gessert\midlotherjointauthor\nametag{$^{1}$}} \Email{nils.gessert@tuhh.de}\\
\Name{Alexander Schlaefer\nametag{$^{1}$}} \Email{schlaefer@tuhh.de}\\
\addr $^{1}$Institute of Medical Technology, Hamburg University of Technology, Germany \\
}

\begin{document}

\maketitle

\begin{abstract}
Autism spectrum disorder (ASD) is associated with behavioral and communication problems. Often, functional magnetic resonance imaging (fMRI) is used to detect and characterize brain changes related to the disorder. Recently, machine learning methods have been employed to reveal new patterns by trying to classify ASD from spatio-temporal fMRI images. Typically, these methods have either focused on temporal or spatial information processing. Instead, we propose a 4D spatio-temporal deep learning approach for ASD classification where we jointly learn from spatial and temporal data. We employ 4D convolutional neural networks and convolutional-recurrent models which outperform a previous approach with an F1-score of $0.71$ compared to an F1-score of $0.65$.

\end{abstract}

\begin{keywords}
4D Deep Learning, 4D CNN, fMRI, ASD
\end{keywords}

\section{Introduction}

Autism spectrum disorder (ASD) is typically associated with difficulties in communication, repetitive behavior and restricted interests. The developmental disorder leads to life-long disability \cite{howlin2004adult}. To understand the underlying causes and find treatments, functional magnetic resonance imaging (fMRI) has been employed which has lead to potential biomarkers of the disease \cite{loth2016identification}. For example, functional connectivity computed from resting-state fMRI has been used to derived features for ASD classification with classical machine learning approaches such as support vector machines \cite{plitt2015functional}. Recently, deep learning approaches have been proposed for ASD classification \cite{wen2018deep}. For example, Dvornek et al. have used long short-term memory (LSTM) cells with sequences extracted from fMRI data \cite{dvornek2017identifying}. The input sequences were computed as the average signal over $200$ manually selected regions of interest for each time point in the 4D fMRI image. Instead of using this handcrafted feature approach, Li et al. \cite{li20182} directly learned spatial features from the fMRI data using 3D convolutional neural networks (CNNs). However, this approach only considered temporal information implicitly by taking the mean and standard deviation volume over a fixed time window. Instead, we propose to directly learn from the full 4D fMRI sequences. This requires 4D spatio-temporal deep learning techniques which are largely unexplored so far. We employ 4D CNNs which process both spatial and temporal dimensions with convolutions. Moreover, we adopt a convolutional-gated recurrent unit 3D CNN (convGRU-CNN3D), which performs temporal processing first with subsequent spatial processing \cite{gessert2018needle}. We compare our methods to the previous approach of using 3D CNNs with mean and standard deviation volumes derived from the 4D image \cite{li20182}. Also, we consider a typical approach from the natural image domain where time steps are stacked in the channel dimension \cite{Pfister2014}. Summarized, we propose 4D spatio-temporal deep learning for ASD classification from 4D fMRI images. We compare to previous, mostly spatial approaches by employing four models with different ways of processing the temporal information.

\section{Methods}

\textbf{Dataset.} We use the publicly available ABIDE dataset \cite{craddock2013neuro} which contains preprocessed resting-state 4D fMRI sequences of patients with ASD and healthy controls. We use a subset of $184$ patients from the NYU collection site. The sequences are preprocessed with the Connectome Computation System pipeline which includes slice timing correction, motion realignment, intensity normalization and rigid registration to a template. Following \cite{dvornek2017identifying}, we employ bandpass filtering (0.01 - 0.1 $\si{\hertz}$) and no global signal regression. Each 4D fMRI image is spatially downsampled to reduce computationally effort such that each fMRI image has a size of $32\times 32\times 32\times 176$ where the first three dimensions are spatial and the last dimension is temporal. We split the dataset into a training, validation and test set with $134$, $30$ and $30$ subjects, respectively. Validation and test set contain an equal number of ASD and control subjects. 

\begin{figure}
\floatconts
  {fig:model}
  {\caption{The four architectures we employ. Note that the images' color dimension is $c=1$.}}
  {\includegraphics[width=1.0\linewidth]{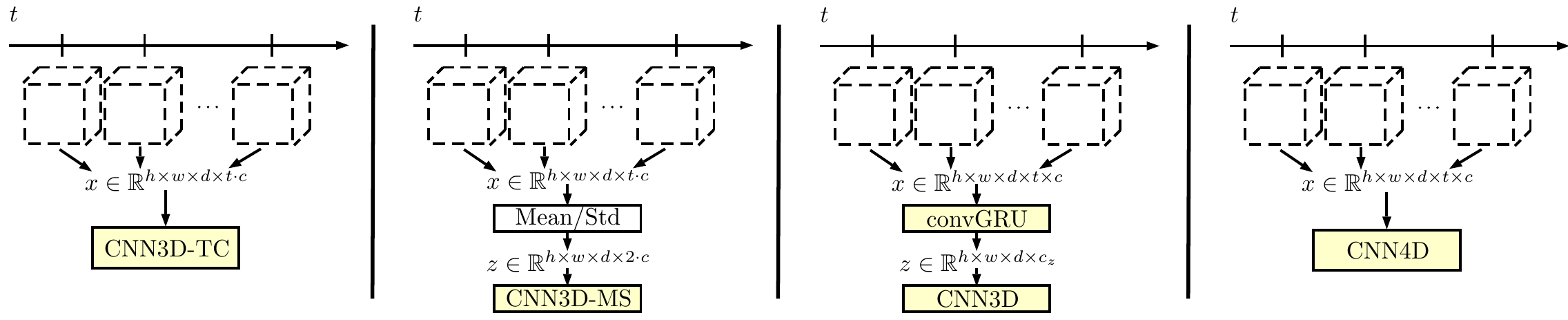}}
\end{figure}

\textbf{Models.} Our deep learning models receive a 4D image of size $32\times32\times32\times15$ as the input, cropped from the full 4D fMRI image. We employ models with four different types of temporal processing, see \figureref{fig:model}. In general, all models follow the idea of densely connected convolutional networks (DenseNet) \cite{Huang2017}. The core architecture is a 3D CNN with one initial convolutional layer, followed by three DenseNet blocks with 5 layers each. 
After the final module we use a global average pooling, which is directly connected to the fully connected layer with two outputs for the two classes. 
\\
For the first type of temporal processing, we treat the CNN's input channel as the temporal dimension \cite{Pfister2014}. Thus, we stack the $15$ volumes into the first layer's input channel (CNN3D-TC). 
\\ 
Second, we compute the voxel-wise mean and standard deviation over the $15$ volumes in a sequence which follows a previous approach for ASD classification \cite{li20182}. Thus, we obtain two 3D images which we stack into the first layer's input channel (CNN3D-MS). 
\\
Third, we learn a volume representation from the $15$ time steps which is fed into the 3D CNN model. For this purpose, we use a gated recurrent neural network with convolutional gating operations (convGRU)  \cite{Xingjian2015} in front of our core DenseNet architecture \cite{gessert2018needle}. The convGRU first performs temporal processing while keeping the data structure intact for subsequent spatial processing by the 3D CNN. The model is trained end-to-end (convGRU-CNN3D). 
\\
Fourth, we employ 4D CNNs by isotropically extending the network's convolutional kernels by a fourth temporal dimension. In contrast to the two-stage setup in convGRU-CNN, the 4D CNN performs spatio-temporal processing throughout the entire network (CNN4D). 


\textbf{Training and Evaluation.} 
During training, we randomly crop sequences of length 15 from the fMRI image of each subject. For validation and testing, we crop and evaluate sequences from fMRI images in a sliding window fashion and average the predictions over all crops of each subject. We train our models for 500 epochs with a batch size of 10. For optimization of the cross-entropy loss function, we employ the Adam optimizer. During training, we evaluate the performance on the validation set every 5 epochs and use the best model for the final evaluation on the test set. 

\section{Results and Discussion}

\begin{table}

\floatconts
  {tab:results}%
  {\caption{Results for all experiments.} \label{tab:All-networks-with metrics}}%
  {\begin{tabular}{lllll}
  & \bfseries CNN3D-TC & \bfseries CNN3D-MS & \bfseries convGRU-CNN3D & \bfseries CNN4D \\ \hline
  Accuracy & $0.57$ & $0.60$ & $\bm{0.67}$ & $0.60$ \\
  F1-score & $0.61$ & $0.65$ & $\bm{0.71}$ & $0.68$ \\ 
\hline
  \end{tabular}}
\end{table}

We report the F1-score and accuracy for our experiments in \tableref{tab:All-networks-with metrics}. In general, the classification accuracy is not very high and in a similar range as previous results on ASD classification with different datasets \cite{dvornek2017identifying}. This underlines that ASD classification from fMRI is a challenging problem. In detail, convGRU-CNN3D shows the highest performance across both metrics, followed by CNN4D. This indicates that explicit learning on the temporal dimension is beneficial for ASD classification from fMRI images. Previously, an approach similar to CNN3D-MS had been used for ASD classification from fMRI images \cite{li20182}. This spatial approach and also CNN3D-TC only implicitly process temporal information. Our more powerful 4D deep learning models with explicit temporal processing appear to be able to learn richer features from the fMRI data. Overall, we propose 4D deep learning models for ASD classification from 4D fMRI sequences. We demonstrate that a 4D spatio-temporal convGRU-CNN3D and CNN4D outperform a previous spatial approach that only implicitly considered temporal information. Our results indicate that 4D deep learning models could be beneficial for other learning problems with 4D fMRI data.

\bibliography{midl-samplebibliography}

\begin{thebibliography}{11}
\providecommand{\natexlab}[1]{#1}
\providecommand{\url}[1]{\texttt{#1}}
\expandafter\ifx\csname urlstyle\endcsname\relax
  \providecommand{\doi}[1]{doi: #1}\else
  \providecommand{\doi}{doi: \begingroup \urlstyle{rm}\Url}\fi

\bibitem[Craddock et~al.(2013)Craddock, Benhajali, Chu, Chouinard, Evans,
  Jakab, Khundrakpam, Lewis, Li, Milham, et~al.]{craddock2013neuro}
Cameron Craddock, Yassine Benhajali, Carlton Chu, Francois Chouinard, Alan
  Evans, Andr{\'a}s Jakab, Budhachandra~Singh Khundrakpam, John~David Lewis,
  Qingyang Li, Michael Milham, et~al.
\newblock {The Neuro Bureau Preprocessing Initiative: open sharing of
  preprocessed neuroimaging data and derivatives}.
\newblock \emph{Neuroinformatics}, 4, 2013.

\bibitem[Dvornek et~al.(2017)Dvornek, Ventola, Pelphrey, and
  Duncan]{dvornek2017identifying}
Nicha~C Dvornek, Pamela Ventola, Kevin~A Pelphrey, and James~S Duncan.
\newblock {Identifying autism from resting-state fMRI using long short-term
  memory networks}.
\newblock In \emph{International Workshop on Machine Learning in Medical
  Imaging}, pages 362--370. Springer, 2017.

\bibitem[Gessert et~al.(2018)Gessert, Priegnitz, Saathoff, Antoni, Meyer,
  Hamann, J{\"u}nemann, Otte, and Schlaefer]{gessert2018needle}
Nils Gessert, Torben Priegnitz, Thore Saathoff, Sven-Thomas Antoni, David
  Meyer, Moritz~Franz Hamann, Klaus-Peter J{\"u}nemann, Christoph Otte, and
  Alexander Schlaefer.
\newblock {Needle Tip Force Estimation Using an OCT Fiber and a Fused
  convGRU-CNN Architecture}.
\newblock In \emph{International Conference on Medical Image Computing and
  Computer-Assisted Intervention}, pages 222--229. Springer, 2018.

\bibitem[Howlin et~al.(2004)Howlin, Goode, Hutton, and Rutter]{howlin2004adult}
Patricia Howlin, Susan Goode, Jane Hutton, and Michael Rutter.
\newblock {Adult outcome for children with autism}.
\newblock \emph{{Journal of child Psychology and Psychiatry}}, 45\penalty0
  (2):\penalty0 212--229, 2004.

\bibitem[Huang et~al.(2017)Huang, Liu, Van Der~Maaten, and
  Weinberger]{Huang2017}
Gao Huang, Zhuang Liu, Laurens Van Der~Maaten, and Kilian~Q. Weinberger.
\newblock {Densely Connected Convolutional Networks}.
\newblock \emph{CVPR}, pages 2261--2269, 2017.

\bibitem[Li et~al.(2018)Li, Dvornek, Papademetris, Zhuang, Staib, Ventola, and
  Duncan]{li20182}
Xiaoxiao Li, Nicha~C Dvornek, Xenophon Papademetris, Juntang Zhuang, Lawrence~H
  Staib, Pamela Ventola, and James~S Duncan.
\newblock {2-channel convolutional 3D deep neural network (2CC3D) for fMRI
  analysis: ASD classification and feature learning}.
\newblock In \emph{ISBI}, pages 1252--1255. IEEE, 2018.

\bibitem[Loth et~al.(2016)Loth, Spooren, Ham, Isaac, Auriche-Benichou,
  Banaschewski, Baron-Cohen, Broich, Boelte, Bourgeron,
  et~al.]{loth2016identification}
Eva Loth, Will Spooren, Lindsay~M Ham, Maria~B Isaac, Caroline
  Auriche-Benichou, Tobias Banaschewski, Simon Baron-Cohen, Karl Broich, Sven
  Boelte, Thomas Bourgeron, et~al.
\newblock {Identification and validation of biomarkers for autism spectrum
  disorders}.
\newblock \emph{Nature Reviews Drug Discovery}, 15\penalty0 (1):\penalty0 70,
  2016.

\bibitem[Pfister et~al.(2014)Pfister, Simonyan, Charles, and
  Zisserman]{Pfister2014}
Tomas Pfister, Karen Simonyan, James Charles, and Andrew Zisserman.
\newblock {Deep convolutional neural networks for efficient pose estimation in
  gesture videos}.
\newblock In \emph{Asian Conference on Computer Vision}, pages 538--552.
  Springer, 2014.

\bibitem[Plitt et~al.(2015)Plitt, Barnes, and Martin]{plitt2015functional}
Mark Plitt, Kelly~Anne Barnes, and Alex Martin.
\newblock {Functional connectivity classification of autism identifies highly
  predictive brain features but falls short of biomarker standards}.
\newblock \emph{NeuroImage: Clinical}, 7:\penalty0 359--366, 2015.

\bibitem[Wen et~al.(2018)Wen, Wei, Zhou, Li, Zhang, and Han]{wen2018deep}
Dong Wen, Zhenhao Wei, Yanhong Zhou, Guolin Li, Xu~Zhang, and Wei Han.
\newblock {Deep learning methods to process fMRI data and their application in
  the diagnosis of cognitive impairment: a brief overview and our opinion}.
\newblock \emph{Frontiers in neuroinformatics}, 12:\penalty0 23, 2018.

\bibitem[Xingjian et~al.(2015)Xingjian, Chen, Wang, Yeung, Wong, and
  Woo]{Xingjian2015}
S.~H.~I. Xingjian, Zhourong Chen, Hao Wang, Dit-Yan Yeung, Wai-Kin Wong, and
  Wang-chun Woo.
\newblock {Convolutional {LSTM} network: A machine learning approach for
  precipitation nowcasting}.
\newblock In \emph{Advances in Neural Information Processing Systems}, pages
  802--810, 2015.

\end{thebibliography}
\end{document}